# Photocarrier relaxation in two-dimensional semiconductors

Daichi Kozawa[a], Rajeev Sharma Kumar[b,c], Alexandra Carvalho[b,c],

Amara Kiran Kumar[b,d], Weijie Zhao[b,c], Shunfeng Wang[b,c], Minglin Toh[b,c],

Ricardo M. Ribeiro[c,e], A. H. Castro Neto[b,c], Kazunari Matsuda[a], Goki Eda[b,c,d,*]

[a]*Institute of Advanced Energy, Kyoto University, Uji, Kyoto, Japan 611-0011*

[b]*Department of Physics, National University of Singapore, 2 Science Drive 3, Singapore 117542*

[c]*Graphene Research Centre, National University of Singapore, 6 Science Drive 2, Singapore 117546*

[d]*Department of Chemistry, National University of Singapore, 3 Science Drive 3, Singapore 117543*

[e]*Center of Physics and Department of Physics, University of Minho, PT-4710- 057, Braga, Portugal*

[*]E-mail: g.eda@nus.edu.sg

## ABSTRACT

Two-dimensional (2D) crystals of semiconducting transition metal dichalcogenides (TMD) absorb a large fraction of incident photons in the visible frequencies despite being atomically thin. It has been suggested that the strong absorption is due to the parallel band or "band nesting" effect and corresponding divergence in the joint density of states. Here, we show using photoluminescence excitation spectroscopy that the band nesting in mono- and bilayer $MX_2$ (M = Mo, W and X = S, Se) results in excitation-dependent characteristic relaxation pathways of the photoexcited carriers.

Our experimental and simulation results reveal that photoexcited electron-hole pairs in the nesting region spontaneously separate in the *k*-space, relaxing towards immediate band extrema with opposite momentum. These effects imply that the loss of photocarriers due to direct exciton recombination is temporarily suppressed for excitation in resonance with band nesting. Our findings highlight the potential for efficient hot carrier collection using these materials as the absorbers in optoelectronic devices.

Quasi-two-dimensional (2D) properties of layered transition metal dichalcogenides (TMDs) have long attracted interest in fundamental research[1]. Since the successful isolation of graphene and other atomically thin crystals from the bulk layered compounds, renewed interest in the TMDs in their ultimate thickness regime led to the discovery of a range of their unique properties[2-4]. Recent findings on the novel optical properties of semiconducting 2D TMDs such as $MoS_2$ and $WSe_2$ include emerging band-gap photoluminescence (PL)[5,6], controlled valley polarization[7-9], strongly bound trions[10,11], second harmonic generation[12,13], strain-induced optical gap modulation[14,15], and surface sensitive luminescence[16-18]. Along with the attractive electrical properties[19] and recent progress in the materials synthesis[20,21], these 2D TMDs show positive prospects for their applications in optoelectronic devices[22,23].

Individual layers of TMDs are excellent light absorbers despite being atomically thin[22,24]. Absorption spectra of $MX_2$ (M = Mo, W and X = S, Se) consist of characteristic peaks due to excitonic resonance and interband transitions. Recent *ab initio* studies have attributed the strong light-matter interaction to the "band nesting", which gives rise to singularity features in the joint density of states (JDOS)[25]. In the nesting region of the band structure, the conduction and valence bands are parallel to each other. The JDOS diverges for the resonance energy, resulting in giant enhancement in the corresponding optical conductivity. While singularities in the JDOS are present in common semiconductors and metals such as silicon and aluminum, divergence in optical conductivity occur only in low-dimensional materials[26]. The absorption is also enhanced for photon energies corresponding to transitions between van Hove singularity peaks in the DOS, which are attributed to heavy effective mass of carriers in $MX_2$ compounds[25,27].

Since absorption is highly efficient in the resonance conditions, understanding photocarrier relaxation dynamics is crucial in implementing these materials into light harvesting devices. Photocarriers generated in the band nesting region are of particular interest as the electrons and holes are expected to relax at the same rate, but with opposite momentum. Here, we examine the relaxation processes of photoexcited carriers in mono- and bilayer $MX_2$ using photoluminescence excitation (PLE) spectroscopy and *ab initio* density functional theory (DFT) calculation. We show that the relaxation channel of the photoexcited carriers is strongly dependent on the excitation energy. Our findings reveal the unique implication of 2D band structure and the characteristic photocarrier relaxation.

Semiconducting group 6 TMDs consist of strongly bonded two-dimensional X-M-X layers that are held together by weak van der Waals forces[1,27]. Each layer of the M and X atoms forms a two-dimensional hexagonal crystal (Fig. 1a). Monolayer $MX_2$ is noncentrosymmetric, while bilayer and bulk materials exhibit inversion symmetry[28]. The band gap is direct for monolayer $MX_2$ in contrast to indirect gap of bilayer and thicker multilayers[5,6]. Since group 6 $MX_2$ is isoelectronic, the general features of the electronic structure are similar. For monolayer $MX_2$, the conduction band minimum (CBM) and the valence band maximum (VBM) are both at the K point of the Brillouin zone (Figure 1b)[29-33]. Conduction band valley at the $\Lambda$ point and the valence band hill at the $\Gamma$ point play a dominant role in the direct to indirect gap crossover. Band nesting regions in Mo and W disulphide and diselenide monolayers occur midway between the $\Gamma$ and $\Lambda$ points[25].

The optical signatures of the $MX_2$ materials reflect basic features of the energy band structure. Figure 1c shows the PL and differential reflectance ($\Delta R/R$) spectra of monolayer $MoS_2$, $MoSe_2$, $WS_2$ and $WSe_2$ on a quartz substrate. The differential reflectance is an effective measure of the absorbance for ultrathin samples[34]. The resonance peaks A and B (labeled according to earlier conventions[33]) correspond to excitonic transitions occurring at the K/K' points of the *k*-space (See Figure 1b)[5,35,36]. Strong absorption at higher energies (C peak for $MoS_2$, $WS_2$, and $MoSe_2$), which is estimated to be above 30% based on reflectance and transmittance measurements, is predicted to be largely due to the band nesting effect[25]. A recent study has shown that this excitation resonance is excitonic in nature[37]. The absorption features of $WSe_2$ are typically more complex due to strong effects of the Se p orbitals[33] but the strong absorption above ~2.4 eV may be attributed partly to band nesting.

The calculated band structure indicates that excited electrons and holes in the nesting region relax to their immediate band extremum. This corresponds to $\Lambda$ valley and $\Gamma$ hill for electrons and holes, respectively. The intraband relaxation is known to be extremely fast (< 500 fs) in mono- and few-layer $MoS_2$ and dominates other relaxation processes[38-40]. Radiative recombination of the electron-hole pairs separated in the *k*-space requires emission or absorption of a phonon and is a slow process, resulting in low yield emission. The carrier lifetime in the indirect emission process can be estimated to be on the order of 1 ns based on previous studies[38,41]. On the other hand, direct excitons exhibit lifetime on the order of 100 ps with correspondingly higher QY[38]. Here, we investigate the effect of band nesting on the behavior of photoexcited electron-hole pairs in monolayer $MX_2$ by studying their excitation-dependent PL quantum yield (QY). To obtain micro-PLE spectra from micrometer-sized flakes of mechanically exfoliated samples, we utilized a supercontinuum light source coupled to a tunable laser filter as the excitation source (See Methods for details). The excitation power was kept constant and low to avoid undesired nonlinear effects.

Figure 2 shows the PLE intensity map of monolayer (a) $MoS_2$, (b) $MoSe_2$, (c) $WS_2$ and (d) $WSe_2$ flakes and comparisons between the PLE spectra (red plots), relative QY of emission (blue dots) and the differential reflectance spectra (gray lines). The relative QY is estimated by dividing the integrated PL intensity by the absorption at the excitation energy. It can be seen that the PLE intensity is enhanced when the excitation is in resonance with the B exciton absorption. In contrast, for $MoS_2$ and $MoSe_2$, the PLE intensity at the C absorption peak is suppressed or only weakly enhanced. We note that similar suppression of the PLE is seen for $WS_2$ and $WSe_2$ for excitation near the C and B' absorption peaks, respectively. In all cases, the relative QY drops consistently for excitation energies above the B peak resonance. We attribute this behavior to the spontaneous separation of electron-hole pairs in the $k$-space due to band nesting.

In order to further verify this effect, we studied the PLE spectra of bilayer $MoS_2$ (Fig. 3a). The absorption features of bilayer samples are similar to those of monolayers, exhibiting excitonic resonance peaks A and B and strong C peak absorption due to band nesting (Fig. 3b). Bilayer $MoS_2$ and $WS_2$ exhibit indirect band gap with a CBM at the $\Lambda$ point and VBM at the $\Gamma$ point[32,34] (Fig. 3c). Thus, we expect enhancement in the band gap indirect emission when the carriers are excited in the band nesting region. The PLE spectrum of the indirect peak (I) shows a distinct enhancement at the C absorption peak (Fig. 3b). On the other hand, the A emission peak shows relatively weaker enhancement in this energy range (Fig. 3b). Figure 3c and d show the evolution of the ratio of indirect (I) and direct (A) peak. Similar results were obtained for bilayer $WS_2$ (See Supplementary Information for details). These results further demonstrate that the photocarrier relaxation pathways and the radiative emission channels are strongly dependent on the excitation energy.

The relaxation pathway of the photocarriers depends strongly on the two-dimensional landscape of the conduction and valence bands in the $k$-space. The carriers relax through phonon scattering, and are subject to the selection rules imposed by energy and momentum conservation. In the approximation of weakly interacting electron and hole, the electron-phonon scattering time is given by the Fermi golden rule

$$\tau^{-1} = \frac{2\pi}{\hbar} \sum_f \left|\langle \psi_i | \hat{V} | \psi_f \rangle\right|^2 \delta(\varepsilon_f - \varepsilon_i \mp \hbar\omega) \left( n_B(\hbar\omega, T) + \begin{Bmatrix} 1 \\ 0 \end{Bmatrix} \right)$$

where $|\psi_i\rangle$ and $|\psi_f\rangle$ are the initial and final states with momentum $k_i$ and $k_f = k_i - q$ and energies $\varepsilon_i$, $\varepsilon_f$, and $\hat{V}$ is the interaction potential, and $n_B(\hbar\omega, T)$ is the Bose-Einstein distribution. The upper/lower elements correspond to phonon emission/absorption.

Figure 4a shows the potential contour of the nesting region in the first Brillouin zone along with the potential map of the conduction and the valence bands. We have defined these regions using the criteria $\left|\nabla_k (E_c - E_v)\right| \ll 1$ eV/($2\pi/a$)[25], where $2\pi/a$ is the modulus of the reciprocal lattice vector, $E_c$ and $E_v$ are the energy in the unoccupied states in the conduction band ($c$) and the occupied states in the valence band (*v*), respectively. This region is highlighted in white in Fig. 4a. The arrows indicate the possible relaxation paths for electrons and holes after excitation in the nesting region. Note that monolayer $WS_2$ exhibits energy landscape similar to that of monolayer $MoS_2$. (See Figure S2). Similar potential maps are shown for bilayer $MoS_2$ (Fig. 4b). It can be seen that the nesting region is extended around the Γ point for bilayer compared to the case of monolayer.

We estimate the fraction of photocarriers reaching the CBM and VBM at the K point after excitation at the nesting region for monolayer $MoS_2$. Both acoustic and optical/homopolar phonons can intervene in the carrier relaxation, the former mainly through deformation potential interaction, and the latter mainly through Fröhlich interaction. In $MoS_2$, the scattering rates have been calculated for both acoustic and optical phonons and were found to be comparable above the onset of optical phonon emission[42]. Further, both rates are approximately independent of *q* except for short wavelength acoustic phonons. Thus, we assume that all allowed phonon emission events are equally probable. The relaxation of the electron and hole were assumed to be independent and the process was stopped whenever the carrier was within a capture radius *R* (taken to be ~$10^{-4}$ $2\pi/a$) of a band extremum.

The fraction of electron-hole pairs that end the relaxation at the K point is shown as a function of the excitation energy in Fig. 4c for monolayer $MoS_2$. Between the energies corresponding to peaks A and B, the probability of electron-hole pairs relaxing to K is unity. However, this fraction becomes considerably lower at ~2 eV. This trend reproduces well the decrease of the relative QY at the C peak (Fig. 2a). From this energy onwards, the relaxation is mostly mediated by acoustic phonons.

When the excitation energy is close to the C peak energy, a large fraction of the holes relax to the Γ point and electrons to the D* point (close to Λ). The energy of the first peak in the optical conductivity is due to nesting and is found at about 2.5 eV as indicated by the arrow in Fig 4c. The population of relaxed carriers at the different stationary points of the Brillouin zone can be found in Supplementary Information.

For bilayers $MoS_2$, a similar trend is observed for the fraction of electron-hole pairs that reach the K point. The band nesting and corresponding divergence in the optical conductivity is observed at excitation energy of ~2.4 eV. At this energy, rapid increase in the population of electrons and holes that relax to the Λ valley and Γ hill is observed. This trend explains the experimentally observed increase in the indirect emission intensity at the C peak absorption in bilayer $MoS_2$ (Fig. 3b).

We summarize in Fig. 5 the possible relaxation pathways for monolayer and bilayer $MX_2$ when the excitation is in resonance with the band nesting energy. The system is initially excited from the ground state (GS) to the band nesting excited state (BN). A large fraction of these excited states relaxes to another excited state (Λ/Γ), which represents the state where electrons occupy the Λ valley and holes are at the Γ hill. Radiative recombination from this state competes unfavorably with the non-radiative decay ($k_{nr}^i$), which is a fast process (2-4 ps)[38], and intervalley scattering ($k_{iv}$) to the lowest excited state (K/K), where both electrons and holes occupy the K point. Thus, only a small fraction of the initial excited states are transferred to the K/K state where radiative decay occurs with a moderate yield. In bilayer $MX_2$, a sizeable fraction of the initial excited states decay to the Λ/Γ state where radiative indirect transition occurs with a modest yield. Note that this indirect emission does not compete with the intervalley scattering unlike the case of monolayer. Hot electron emission from the K/K state also occurs with non-negligible efficiency, partly due to intraband relaxation and intervalley scattering of hot carriers.

The above qualitative model indicates that nonradiative decay rate $k_{nr}$ plays a crucial role in the carrier relaxation channels and emission intensity. It is known that $k_{nr}$ depends on the density of nonradiative decay centers, which may come from the trap states of the substrate, surface impurities, and intrinsic defects of the material[38,41]. We investigated the effects of nonradiative decay channels by arbitrarily reducing $k_{nr}$ by using the hexagonal boron nitride as the substrate. We found that this leads not only to the overall enhancement in the PL intensity but also slight increase in the relative QY at the C absorption peak, supporting the validity of the above model (See Supplementary Information for details).

Strong light-matter interaction in semiconducting 2D TMDs is a remarkable feature that makes these materials attractive for their use in optoelectronics. Band nesting plays a crucial role in the spontaneous separation of electron-hole pairs in the *k*-space and temporary suppression of their relaxation to the fundamental band edge. Our findings provide insight into the unique dynamics of photocarrier relaxation pathways and motivate studies on the hot electron behaviors in 2D materials and their potential for efficient hot carrier collection devices.

**Methods**

**Sample preparation**

The samples used in this study were monolayer (and bilayer) $MoS_2$, $MoSe_2$, $WS_2$ and $WSe_2$ crystalline flakes. For $MoS_2$, we studied mechanically exfoliated flakes as well as chemical vapor deposition (CVD)-grown samples (See Supplementary Information for details). Monolayer $MoS_2$ on quartz substrate was directly grown by CVD. Single bulk crystals of $MoSe_2$, $WS_2$ and $WSe_2$ were grown by chemical vapor transport using iodine as the transport agent. The crystals were mechanically exfoliated and deposited on quartz substrates. The number of layers was verified by PL, differential reflectance and Raman spectra.

**Optical measurements.**

The measurements of differential reflectance were performed using a tungsten-halogen lamp. The micro-PL spectra under a back scattering geometry were obtained by monochromer and a super-continuum light as an excitation source coupled to a tunable laser filter. The excitation intensities for PL and PLE measurements were kept below 10 μW in which the effect of optical non-linearity is negligible (not shown). The measured spectral data were corrected for variations in the detection sensitivity with the correction factors obtained by using a standard tungsten-halogen lamp. The differential reflectance is defined as $(R_{S+Q} - R_Q)/R_Q$ where $R_{S+Q}$ and $R_Q$ are the reflected light intensities from the quartz substrate with and without the material, respectively.[6,34,43]

**Theoretical calculations.**

We performed a series of DFT calculations for the semiconductor TMDs family using the open source code QUANTUM ESPRESSO[44]. We used norm-conserving, fully relativistic pseudopotentials with nonlinear core-correction and spin-orbit information to describe the ion cores[45]. The exchange correlation energy was described by the generalized gradient approximation (GGA), in the scheme proposed by Perdew, Burke, and Ernzerhof (PBE)[46]. The integrations over the Brillouin zone were performed using a scheme proposed by Monkhorst-Pack[47,48] for all calculations. We calculated the optical conductivity directly from the band structure[49]. It is well known that GGA underestimates the band-gap[29], and hence the optical conductivity shows the peaks displaced towards lower energies relative to actual experiments. However, their shapes and intensities are expected to be correct.

**References**

1   Wilson, J. A. & Yoffe, A. D. Transition metal dichalcogenides discussion and interpretation of observed optical, electrical and structural properties. *Adv. Phys.* **18**, 193-335 (1969).

2   Wang, Q. H., Kalantar-Zadeh, K., Kis, A., Coleman, J. N. & Strano, M. S. Electronics and optoelectronics of two-dimensional transition metal dichalcogenides. *Nat. Nanotechnol.* **7**, 699-712 (2012).

3   Li, X., Zhang, F. & Niu, Q. Unconventional quantum Hall effect and tunable spin Hall effect in Dirac materials: Application to an isolated $MoS_2$ trilayer. *Phys. Rev. Lett.* **110**, 066803 (2013).

4   Eda, G. *et al.* Coherent atomic and electronic heterostructures of single-layer $MoS_2$. *ACS Nano* **6**, 7311-7317 (2012).

5   Splendiani, A. *et al.* Emerging photoluminescence in monolayer $MoS_2$. *Nano Lett.* **10**, 1271-1275 (2010).

6   Mak, K. F., Lee, C., Hone, J., Shan, J. & Heinz, T. F. Atomically thin $MoS_2$: A new direct-gap semiconductor. *Phys. Rev. Lett.* **105**, 136805 (2010).

7   Mak, K. F., He, K., Shan, J. & Heinz, T. F. Control of valley polarization in monolayer $MoS_2$ by optical helicity. *Nat. Nanotechnol.* **7**, 494-498 (2012).

8   Xiao, D., Liu, G.-B., Feng, W., Xu, X. & Yao, W. Coupled spin and valley physics in monolayers of $MoS_2$ and other group-VI dichalcogenides. *Phys. Rev. Lett.* **108**, 196802 (2012).

9   Cao, T. *et al.* Valley-selective circular dichroism of monolayer molybdenum disulphide. *Nat. Commun.* **3**, 887 (2012).

10  Mak, K. F. *et al.* Tightly bound trions in monolayer $MoS_2$. *Nat. Mater.* **12**, 207-211 (2013).

11  Ross, J. S. *et al.* Electrical control of neutral and charged excitons in a monolayer semiconductor. *Nat. Commun.* **4**, 1474 (2013).

12  Kumar, N. *et al.* Second harmonic microscopy of monolayer $MoS_2$. *Phys. Rev. B* **87**, 161403(R) (2013).

13  Zeng, H. L. *et al.* Optical signature of symmetry variations and spin-valley coupling in atomically thin tungsten dichalcogenides. *Sci. Rep.* **3**, 1608 (2013).

14  He, K., Poole, C., Mak, K. F. & Shan, J. Experimental demonstration of continuous electronic structure tuning via strain in atomically thin MoS$_2$. *Nano Lett.* **13**, 2931-2936 (2013).

15  Feng, J., Qian, X. F., Huang, C. W. & Li, J. Strain-engineered artificial atom as a broad-spectrum solar energy funnel. *Nat. Photonics* **6**, 865-871 (2012).

16  Mouri, S., Miyauchi, Y. & Matsuda, K. Tunable photoluminescence of monolayer MoS$_2$ via chemical doping. *Nano Lett.* **13**, 5944-5948 (2013).

17  Tongay, S. *et al.* Defects activated photoluminescence in two-dimensional semiconductors: interplay between bound, charged, and free excitons. *Sci. Rep.* **3**, 2657 (2013).

18  Tongay, S. *et al.* Broad-range modulation of light emission in two-dimensional semiconductors by molecular physisorption gating. *Nano Lett.* **13**, 2831-2836 (2013).

19  Radisavljevic, B., Radenovic, A., Brivio, J., Giacometti, V. & Kis, A. Single-layer MoS$_2$ transistors. *Nat. Nanotechnol.* **6**, 147-150 (2011).

20  Lee, Y.-H. *et al.* Synthesis of large-area MoS$_2$ atomic layers with chemical vapor deposition. *Adv. Mater.* **24**, 2320-2325 (2012).

21  Eda, G. *et al.* Photoluminescence from chemically exfoliated MoS$_2$. *Nano Lett.* **11**, 5111-5116 (2011).

22  Britnell, L. *et al.* Strong light-matter interactions in heterostructures of atomically thin films. *Science* **340**, 1311-1314 (2013).

23  Eda, G. & Maier, S. A. Two-dimensional crystals: managing light for optoelectronics. *ACS Nano* **7**, 5660-5665 (2013).

24  Bernardi, M., Palummo, M. & Grossman, J. C. Extraordinary sunlight absorption and one nanometer thick photovoltaics using two-dimensional monolayer materials. *Nano Lett.* **13**, 3664-3670 (2013).

25  Carvalho, A., Ribeiro, R. M. & Castro Neto, A. H. Band nesting and the optical response of two-dimensional semiconducting transition metal dichalcogenides. *Phys. Rev. B* **88**, 115205 (2013).

26  Bassani, G. F. & Parravicini, G. P. *Electronic states and optical transitions in solids*. (Pergamon Press, 1975).

27  Mattheis. L. Band structures of transition-meal-dichalcogenide layer compounds. *Phys. Rev. B* **8**, 3719-3740 (1973).

28      Li, Y. L. *et al.* Probing symmetry properties of few-layer $MoS_2$ and h-BN by optical second-harmonic generation. *Nano Lett.* **13**, 3329-3333 (2013).

29      Komsa, H. & Krasheninnikov, A. Effects of confinement and environment on the electronic structure and exciton binding energy of $MoS_2$ from first principles. *Phys. Rev. B* **86**, 241201(R) (2012).

30      Shi, H., Pan, H., Zhang, Y.-W. & Yakobson, B. I. Quasiparticle band structures and optical properties of strained monolayer $MoS_2$ and $WS_2$. *Phys. Rev. B* **87**, 155304 (2013).

31      Jiang, H. Electronic Band structures of molybdenum and tungsten dichalcogenides by the GW Approach. *J. Phys. Chem. C* **116**, 7664-7671 (2012).

32      Zhao, W. *et al.* Origin of indirect optical transitions in few-layer $MoS_2$, $WS_2$ and $WSe_2$. *Nano Lett.* (2013).

33      Beal, A., Knights, J. & Liang, W. Transmission spectra of some transition metal dichalcogenides. II. Group VIA: Trigonal prismatic coordination. *J. Phys. C: Sol. Stat. Phys.* **5**, 3540-3551 (1972).

34      Zhao, W. *et al.* Evolution of electronic structure in atomically thin sheets of $WS_2$ and $WSe_2$. *ACS Nano* **7**, 791-797 (2013).

35      Coehoorn, R., Haas, C. & Degroot, R. A. Electronic structure of $MoSe_2$, $MoS_2$ and $WSe_2$. II. The nature of the optical band-gaps. *Phys. Rev. B* **35**, 6203-6206 (1987).

36      Zhu, Z. Y., Cheng, Y. C. & Schwingenschloegl, U. Giant spin-orbit-induced spin splitting in two-dimensional transition-metal dichalcogenide semiconductors. *Phys. Rev. B* **84**, 153402 (2011).

37      Qiu, D. Y., da Jornada, F. H. & Louie, S. G. Optical spectrum of $MoS_2$: many-body effects and diversity of exciton states. *Phys. Rev. Lett.* **111**, 216805 (2013).

38      Shi, H. *et al.* Exciton dynamics in suspended mono layer and few-layer $MoS_2$ 2D crystals. *ACS Nano* **7**, 1072-1080 (2013).

39      Sim, S. *et al.* Exciton dynamics in atomically thin $MoS_2$: Interexcitonic interaction and broadening kinetics. *Phys. Rev. B* **88**, 075434 (2013).

40      Wang, R. *et al.* Ultrafast and spatially resolved studies of charge carriers in atomically thin molybdenum disulfide. *Phys. Rev. B* **86**, 045406 (2012).

41   Kumar, N., He, J., He, D., Wang, Y. & Zhao, H. Charge carrier dynamics in bulk MoS$_2$ crystal studied by transient absorption microscopy. *J. Appl. Phys.* **113**, 133702 (2013).

42   Kaasbjerg, K., Thygesen, K. S. & Jacobsen, K. W. Phonon-limited mobility in n-type single-layer MoS$_2$ from first principles. *Phys. Rev. B* **85**, 115317 (2012).

43   Mak, K. F. *et al.* Measurement of the optical conductivity of graphene. *Phys. Rev. Lett.* **101**, 196405 (2008).

44   Giannozzi, P. *et al.* QUANTUM ESPRESSO: a modular and open-source software project for quantum simulations of materials. *J. Phys.: Condens. Matter* **21**, 395502 (2009).

45   The pseudopotentials used were either obtained from the QUANTUM ESPRESSO distribution or produced using the ATOMIC code by A. Dal Corso.

46   Perdew, J., Burke, K. & Ernzerhof, M. Generalized gradient approximation made simple. *Phys. Rev. Lett.* **77**, 3865-3868 (1996).

47   Monkhorst, H. & Pack, J. Special points for brillouin-zone integrations. *Phys. Rev. B* **13**, 5188-5192 (1976).

48   Single-layer samples were modeled in a slab geometry by including a vacuum region of 45 bohr in the direction perpendicular to the surface. A grid of 16×16×1 *k* points was used to sample the BZ. The energy cutoff was 50 Ry. The atomic positions were optimized using the Broyden-Fletcher-Goldfarb-Shanno (BFGS) method for the symmetric structure. The lattice parameter *a* was determined by minimization of the total energy. A Gaussian broadening of 0.05-eV width was applied in the optical conductivity.

49   The joint density of states, the dielectric permittivity, and the optical conductivity were calculated using a modified version of the epsilon program of the quantum espresso distribution (Ref. 44) to account for full relativistic calculations.

**Acknowledgement**

We thank Y. Miyauchi and S. Mouri for many helpful discussions. G.E acknowledges Singapore National Research Foundation for funding the research under NRF Research Fellowship (NRF-NRFF2011-02) and Graphene Research Centre. K.M is thankful for the financial support by a Grant-in-Aid for Scientific Research from MEXT of Japan (Nos. 40311435 and 23340085). R.M.R is thankful for the financial support by FEDER through the COMPETE Program, by the Portuguese Foundation for Science and Technology (FCT) in the framework of the Strategic Project PEST-C/FIS/UI607/2011 and grant nr. SFRH/BSAB/1249/2012 and by the EC under Graphene Flagship (contract no. CNECT-ICT-604391).

**FIGURE CAPTIONS**

**Figure 1 | Atomic structure, electronic band structure, and optical spectra of monolayer MX$_2$. a,** Lattice structures of monolayer and bilayer MX$_2$. **b,** Simplified band structure of monolayer of MX$_2$. The arrows indicate the transition in A, B and the band nesting (BN). **c,** PL spectra (red, green, blue and purple curves) from excitation at the C (A' for WSe$_2$) peak and differential reflectance spectra (gray curves) of monolayer MX$_2$ flakes on quartz substrates. The scale bar indicates 20% absorption based on differential reflectance spectra. The PL intensity is normalized by the A exciton peak of the differential reflectance spectra for each material and the spectra are displaced along the vertical axis for clarity.

**Figure 2 | PLE spectra of monolayer MX$_2$.** PLE intensity map (left panel), PLE spectra and relative quantum yield (QY) of emission (right panel) for band gap emission for monolayer (**a**) MoS$_2$, (**b**) MoSe$_2$, (**c**) WS$_2$ and (**d**) WSe$_2$ flakes. Differential reflectance spectra are also shown for comparison. The PLE spectra are based on the integrated intensity of the A peak in the PL spectra at each excitation energy. Each PLE spectrum is normalized by the B exciton peak of each material.

**Figure 3 | PLE spectra of bilayer MoS$_2$. a, b,** PLE intensity map and PLE spectra for bilayer MoS$_2$. The PLE spectra are based on the integrated intensity of the A and I peak in the PL spectra at each excitation energy. The differential reflectance spectrum is also shown for comparison. The PLE spectrum of the A peak is normalized by the B exciton peak of the differential reflectance and the PLE of the I peak is multiplied by the same factor as the PLE spectrum of the A peak. **c,** PL

spectra collected with excitation energy of 2.02, 2.38, 2.48, 2.76 eV. **d,** The ratio of integrated PL intensity between the I and A peak as a function of excited energy. The differential reflectance spectrum is also shown for comparison.

**Figure 4 | Calculated energy landscape and optical spectra for monolayer and bilayer MoS$_2$. a, b,** Energy map of $E_c$-$E_v$, $E_c$ and $E_v$ in the Brillouin zone for monolayer and bilayer MoS$_2$. The arrows in $E_c$ and $E_v$ indicate possible relaxation pathways of carriers from the nesting region. **c, d,** The fraction of electron-hole pairs that end the relaxation at the K point (P$_{K-K}$, red curve) and the optical conductivity (σ, black curve) for monolayer and bilayer MoS$_2$. For **d**, the fraction of electron-hole pairs relaxing to Λ valley and Γ hill (P$_{Λ-Γ}$) is also shown (blue plot). The arrows in **c** and **d** indicate the position of the first peak due to band nesting.

**Figure 5 | Excitation and relaxation pathways for photocarriers.** Energy diagram representing photocarrier relaxation channels in monolayer and bilayer MX$_2$ where the initial excitation is from the ground state (GS) to the band nesting (BN) energy. Nonraditive transition is indicated with a solid arrow. A rate constant *k* is associated with each transition. The subscripts indicate the types of transition: intravalley thermalization (*th*), intervalley scattering (*iv*), radiative (*r*) and nonradiative (*nr*). The superscripts (*i*) and (*d*) indicate indirect and direct transition, respectively.

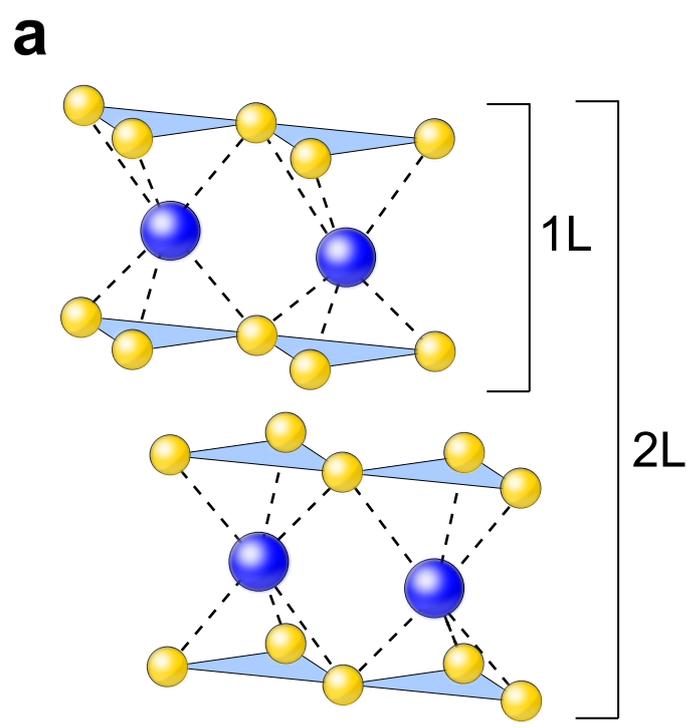
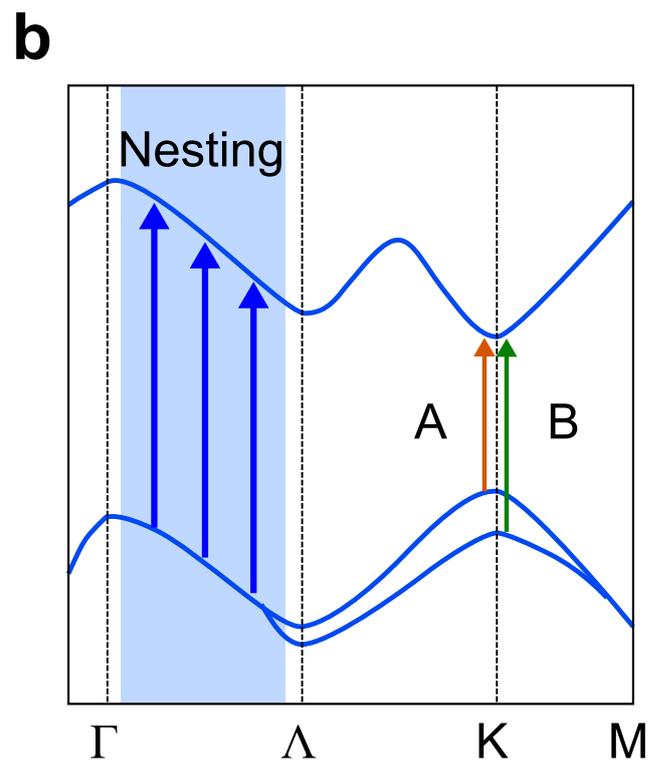
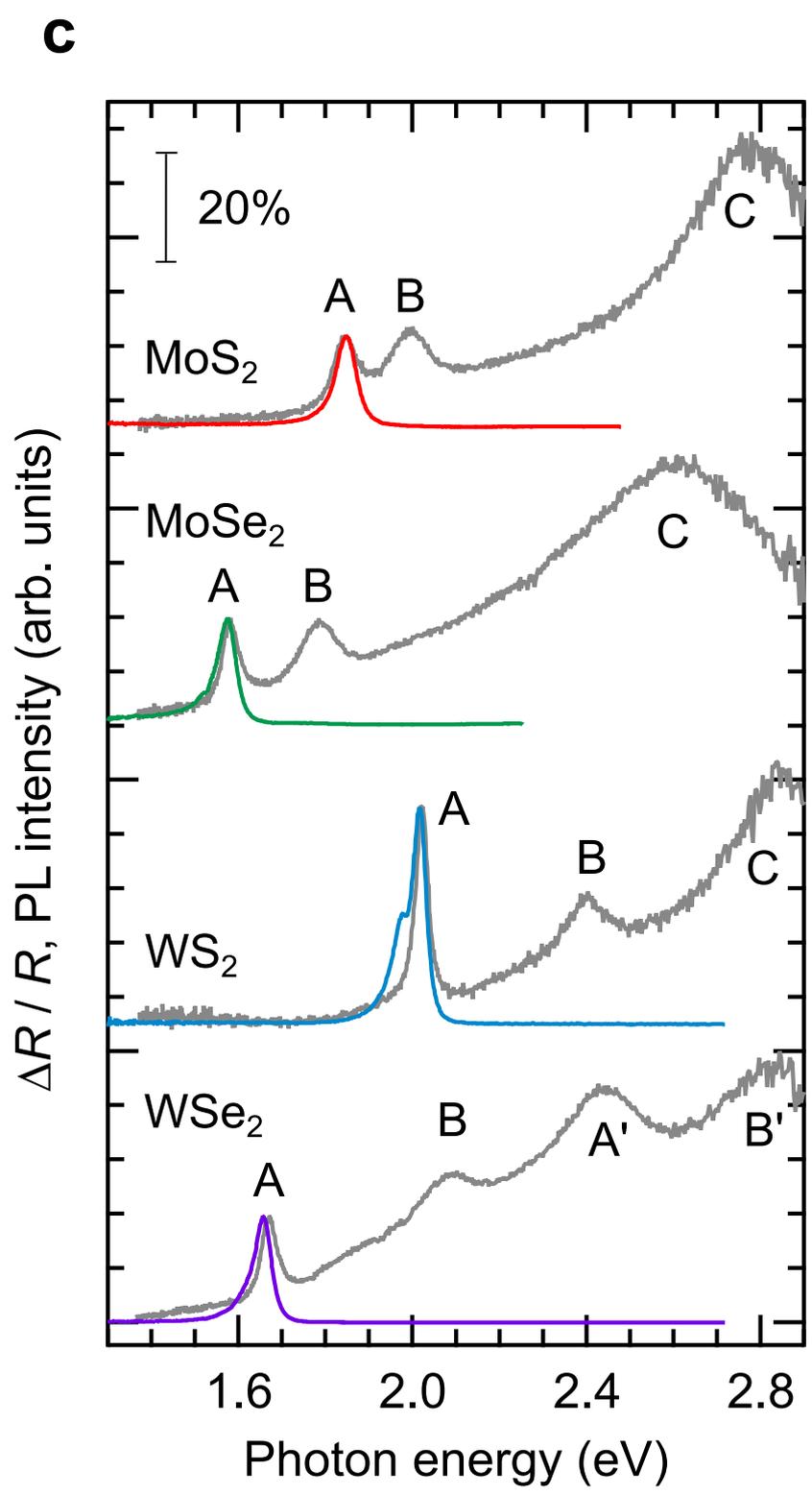

Fig. 1 D. Kozawa *et al.*

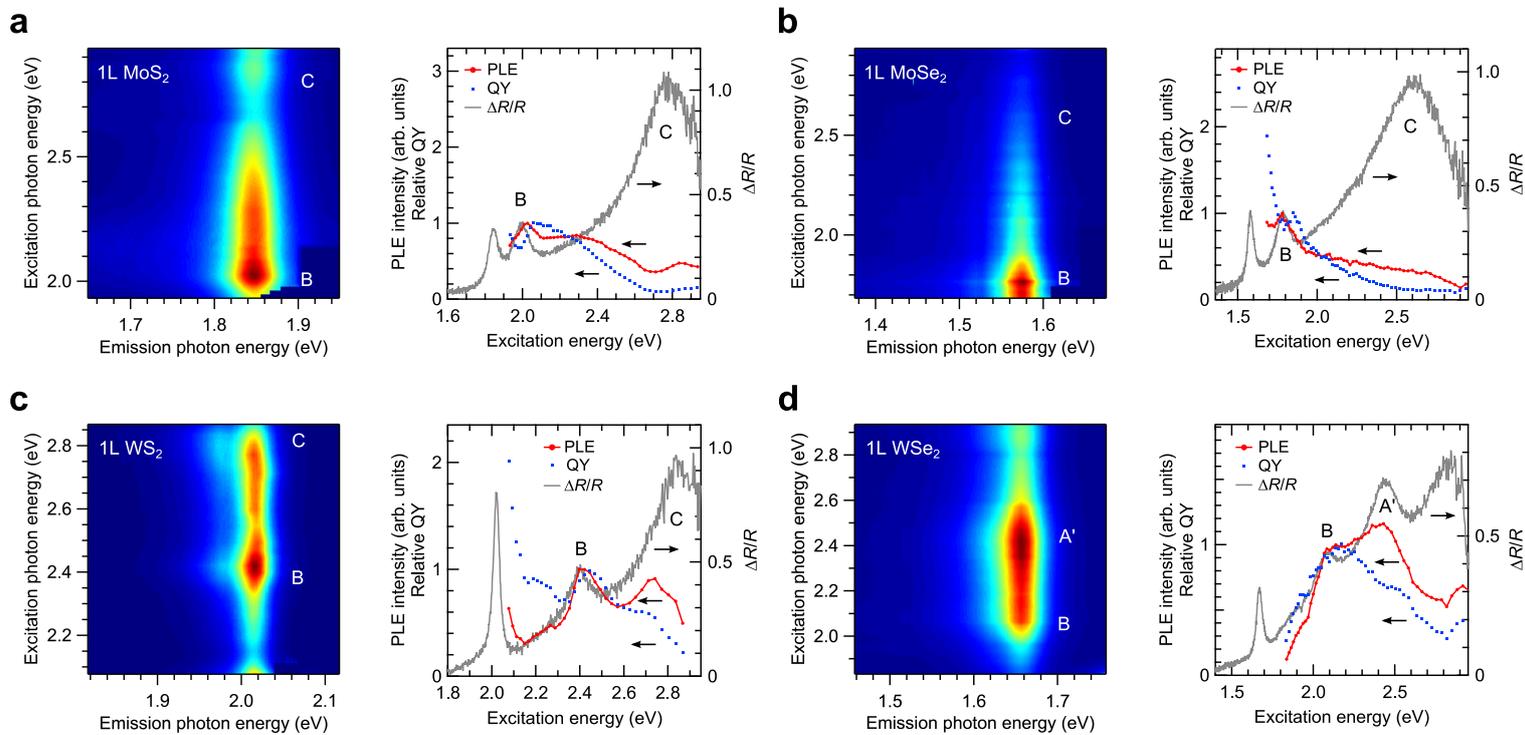

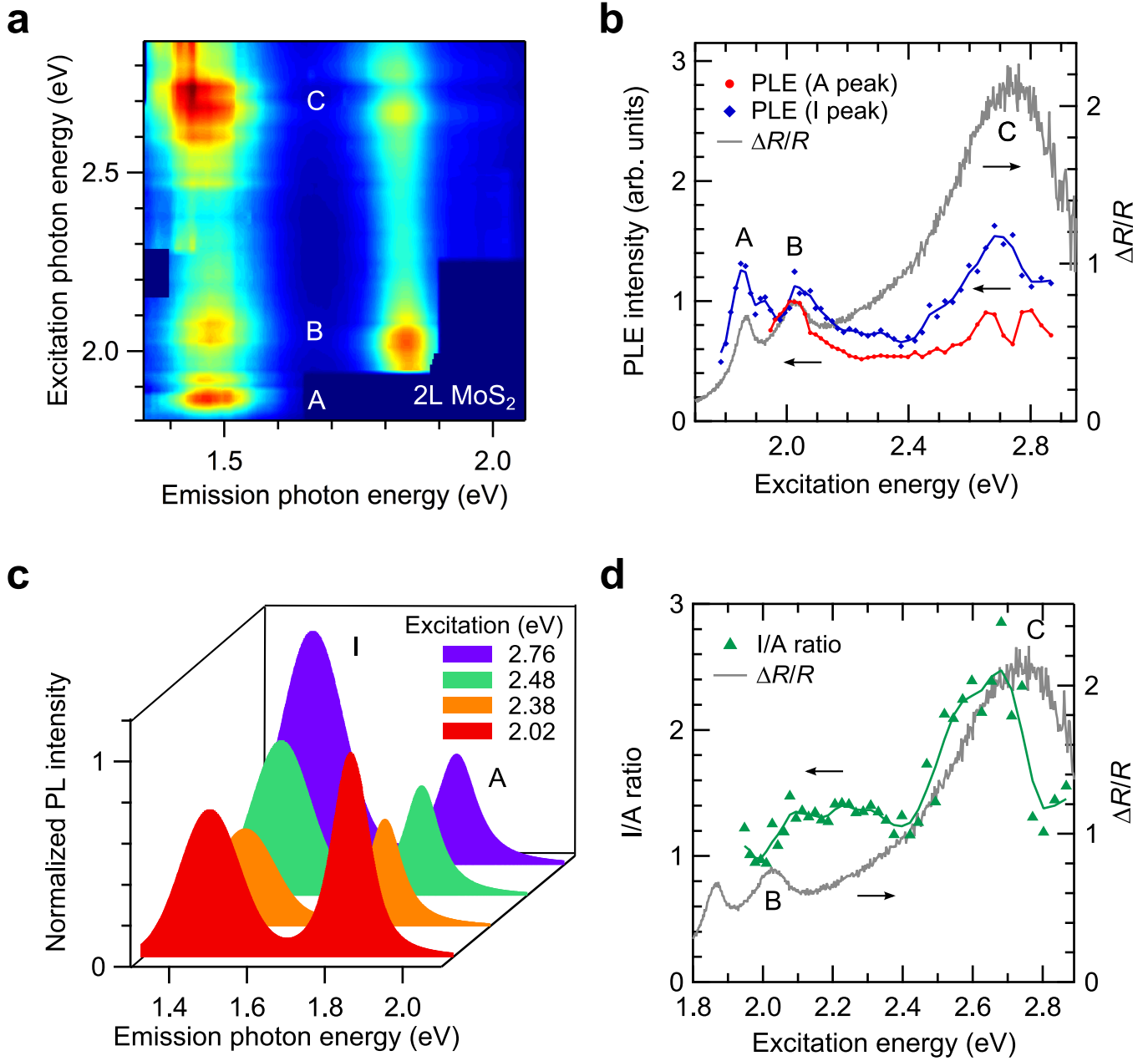

Fig. 3    D. Kozawa *et al.*

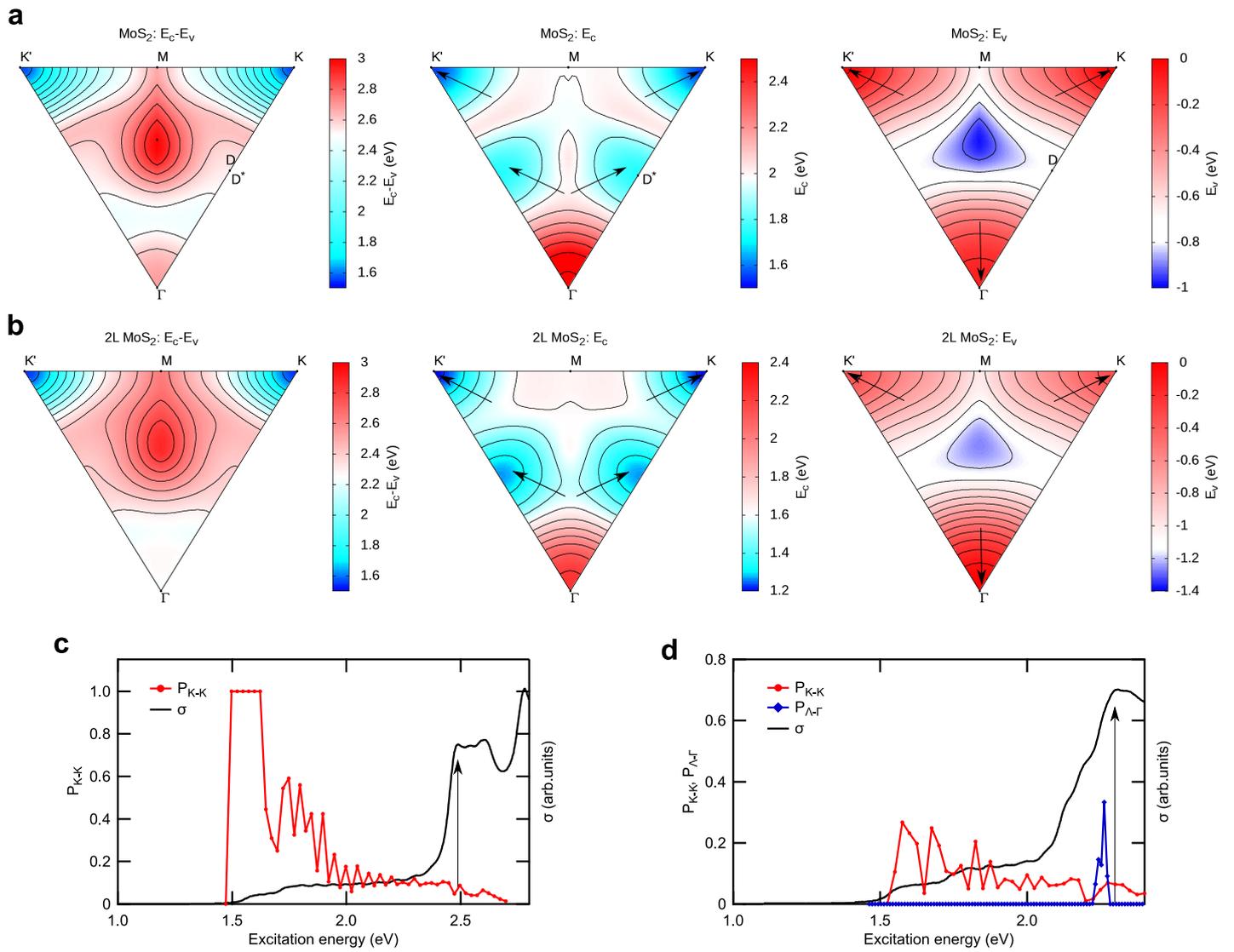

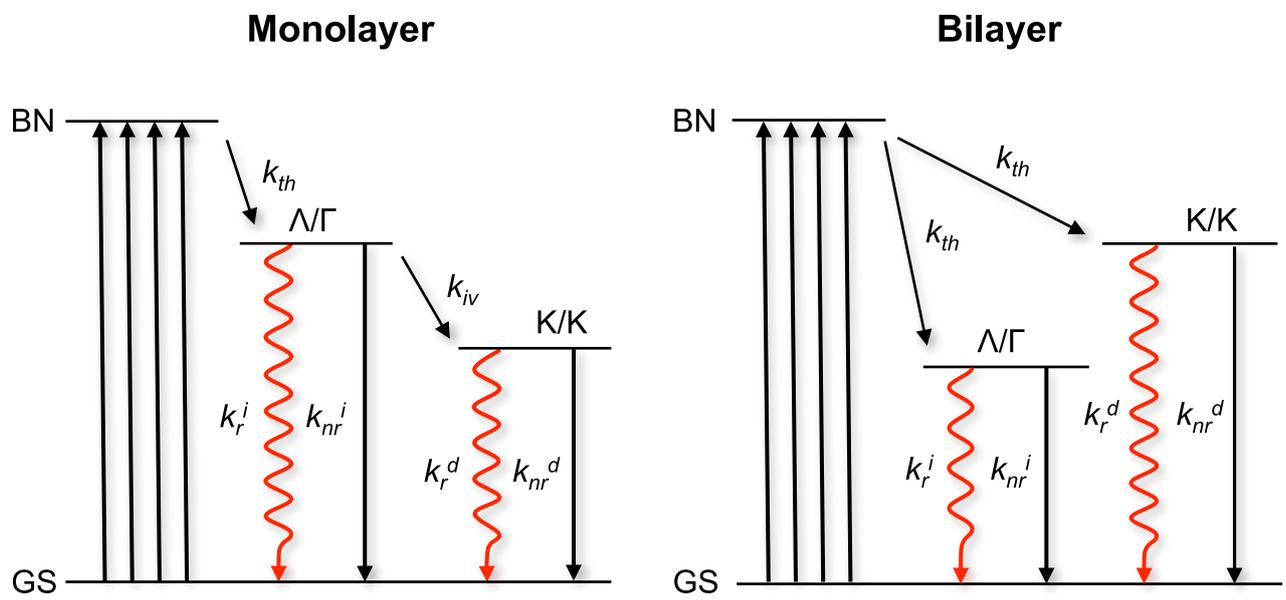

Fig. 5 D. Kozawa *et al.*

# Supplementary Information

# Photocarrier relaxation in two-dimensional semiconductors

Daichi Kozawa[a], Rajeev Sharma Kumar[b,c], Alexandra Carvalho[b,c],

Amara Kiran Kumar[b,d], Weijie Zhao[b,c], Shunfeng Wang[b,c], Minglin Toh[b,c],

Ricardo M. Ribeiro[c,e], A. H. Castro Neto[b,c], Kazunari Matsuda[a], Goki Eda[b,c,d]

[a]*Institute of Advanced Energy, Kyoto University, Uji, Kyoto, Japan 611-0011*

[b]*Department of Physics, National University of Singapore, 2 Science Drive 3, Singapore 117542*

[c]*Graphene Research Centre, National University of Singapore, 6 Science Drive 2, Singapore 117546*

[d]*Department of Chemistry, National University of Singapore, 3 Science Drive 3, Singapore 117543*

[e]*Center of Physics and Department of Physics, University of Minho, PT-4710- 057, Braga, Portugal*

**S1. Photoluminescence Excitation properties of exfoliated monolayer $MoS_2$**

Monolayer $MoS_2$ flakes obtained by various preparation methods were studied to compare the relaxation process of photoexcited carriers. We measured photoluminescence excitation (PLE) for mechanically exfoliated flakes of $MoS_2$ on quartz and $MoS_2$ on hexagonal boron nitride (hBN) substrates as well as chemical vapor deposition (CVD)-Grown samples. Figure S1a displays PL spectra for the monolayer CVD-grown, exfoliated $MoS_2$ and $MoS_2$/hBN. The spectra of exfoliated $MoS_2$ on quartz and hBN substrates are muliplied by a factor of 40 and 2.5 for comparison. Relatively low energy and broadening of PL for exfoliated MoS$_2$ suggests higher doping level and showing charged exciton (trion) recombination[1-4]. Relatively small intensity can be attributed to low decay rate of radiative recombination. MoS$_2$ deposited on hBN is expected to show strong PL due to change in the doping level and reduced non-radiative decay channels due to substrate defects[5,6]. Figure S1 shows PLE map, spectra, and relative quantum yield of emission for exfoliated monolayer MoS$_2$ on (b) quartz and (c) hBN substrates. In either case, weak enhancement in the emission intensity at the C peak excitation was observed, is similar to the case of CVD-grown MoS$_2$ discussed in the main text. This result suggests that the effect of band nesting is intrinsic to the material. Some differences are, however, observed in the ratio of the emission quantum yield for B peak and C peak excitation. This is possibly due to changes in the non-radiative recombination rate $k_{nr}^i$ and intervalley scattering rate $k_{iv}$ as discussed in the main text.

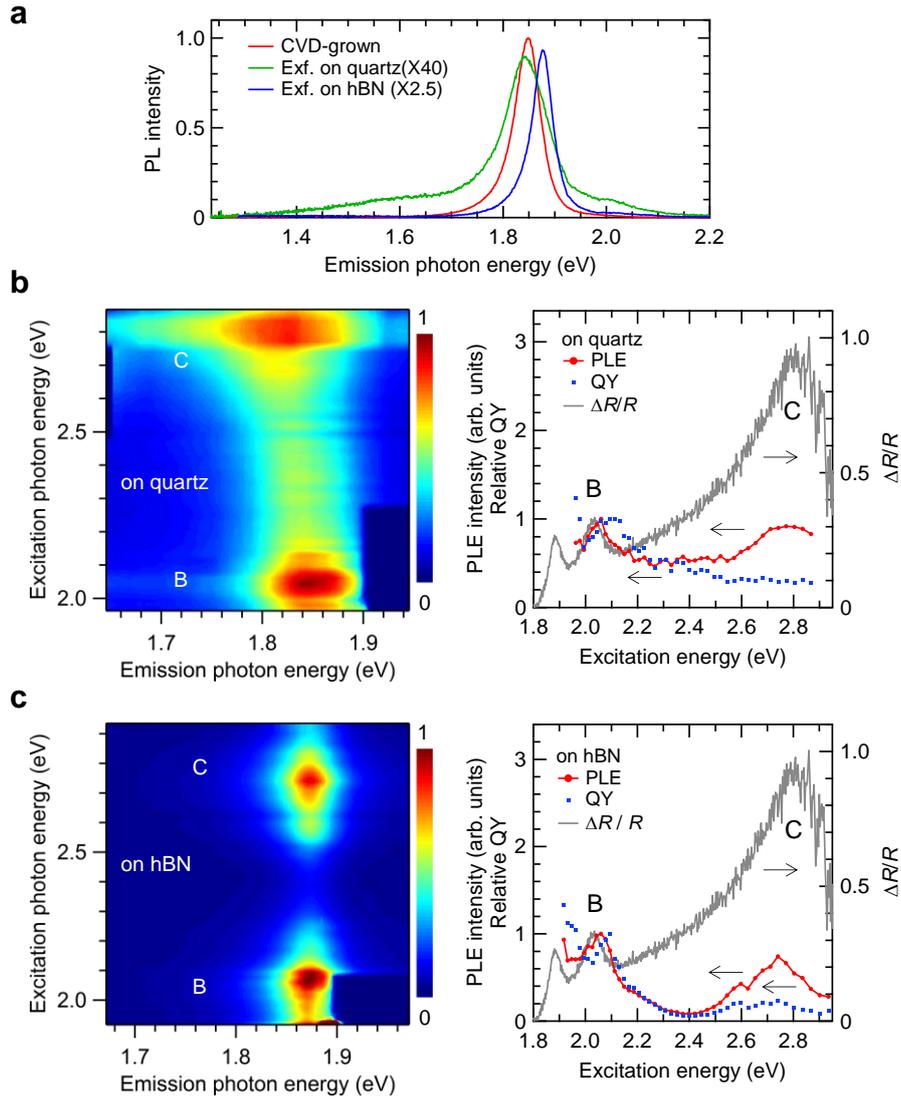

**Figure S1: PL and PLE spectra of monolayer MoS$_2$. a,** PL spectra for monolayer CVD-grown, exfoliated MoS$_2$ on quartz and hBN substrates. PLE intensity map (left panel), PLE spectra and relative quantum yield (QY) (right panel) for band gap emission of exfoliated monolayer on (**b**) quartz and (**c**) hBN substrates.

## S2. Monte-Carlo simulation

The Monte-Carlo simulations of the carrier relaxation paths were carried out using as input the topology of the two highest occupied bands and two lowest unoccupied bands calculated from first-principles (see Fig. 3 and Supplementary Fig. S2), as described in the Methods section. Excitonic effects were not taken into account. The phonon energies were obtained from Ref. 7. The phonon dispersion for the acoustic bands is considered to be approximately linear, whereas the optical phonon energy is considered to be independent on the phonon wavelength. The scattering probability is assumed to be the same for acoustic and optical phonons.[7] The relaxation is stopped when the carrier is within a capture radius $R$ (taken to be $\sim 10^{-4} 2\pi/a$) of a minimum of $E_c$, maximum of $E_v$ or when there are no allowed phonon-emission transitions. The k-point grid interval was taken to be $0.01\ 2\pi/a$, where $a$ is the lattice constant.

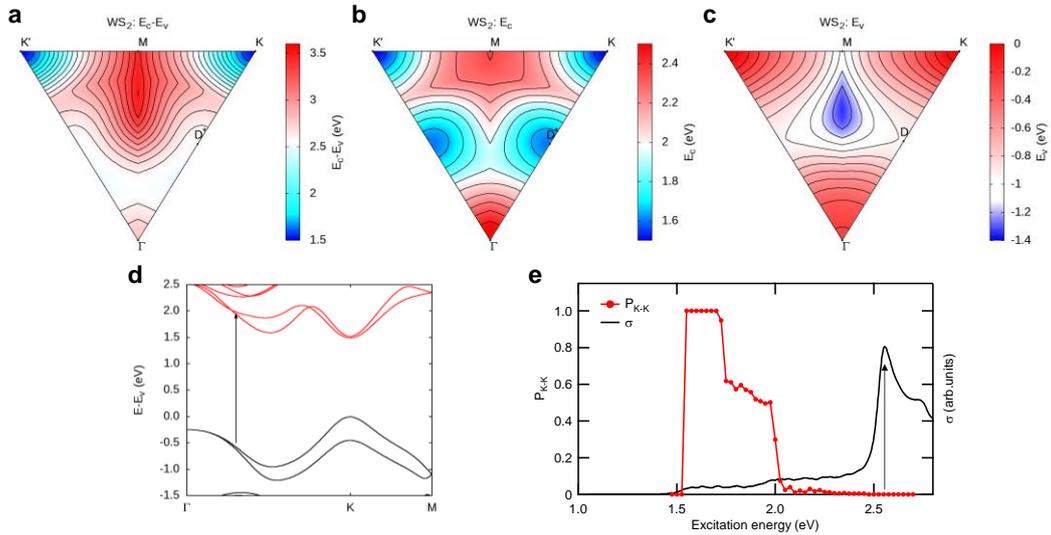

**Supplementary Figure S2: Calculated topology of the electron energy bands for monolayer WS$_2$.** Map on the Brillouin zone of $E_c$-$E_v$ (**a**), $E_c$ (**b**) and $E_v$ (**c**) for monolayer WS$_2$. **d,** the band structure of monolayer WS$_2$. **e,** The fraction of electron-hole pairs that end the relaxation at the K point (red curve, left axis) and the optical conductivity (black curve, right axis). The arrows in **d** and **e** signal the position of the first band-nesting peak.

The Supplementary Figure S3 highlights the difference between the excitation with photons with an energy slightly above B, which in the calculations is located at about 1.7 eV for $MoS_2$, and an energy in the band-nesting region (close to the energy of the C peak). Firstly, it should be noticed that a larger area of the reciprocal space is excited in the second case (with energy 2.50 eV), for the same energy interval of 25 meV – thus the greatest optical conductivity. Photons with energy close to 1.90 eV produce electron-hole pairs in the vicinity of the K and K' points. Those electrons can only relax to the bottom of the conduction band of K, and the respective electrons can relax to K/K' or $\Gamma$ (where $E_v$, according to our calculation, is only about 80 meV below K), and therefore the subsequent intensity of the PL A transition is high. In contrast, for an excitation energy of 2.50 eV, a large fraction of the photo-excited electron-hole pairs is originally close to D or $\Gamma$ than to K or K'. Thus, it is likely that a sizable fraction of the electron-hole pairs is dissociated and later recombines through the indirect transition D*- $\Gamma$.

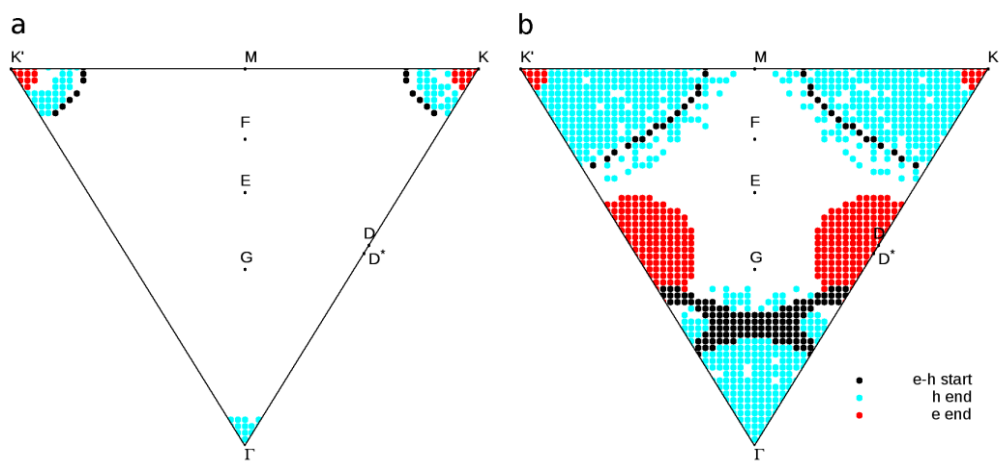

**Figure S3: Calculated end points of the relaxation paths of photoexcited carriers for monolayer $MoS_2$,** for excitation energies (**a**) 1.900±0.025 eV

and (**b**) 2.500±0.025 eV. The plot area represents 1/6th of the Brillouin zone. The K points where electron-hole pairs are generated for that particular excitation energy are marked in black (e-h start). The K points where the electrons and holes ended up after relaxation are shown in color.

## S3. PLE properties of bilayer MoSe$_2$ and WS$_2$

To further verify the spontaneous separation of electron-hole pairs in the phase space, we further studied the PLE spectra of bilayer WS$_2$ (Fig. S4). The PLE spectrum of the indirect peak (I) shows a distinct enhancement at the C absorption peak. Similar trend is seen in bilayer MoS$_2$ (Fig. 3b). The A emission peak also shows enhancement in this energy range. The enhancement is probably contributed by excitation of electron to the second lowest conduction band at the K points.

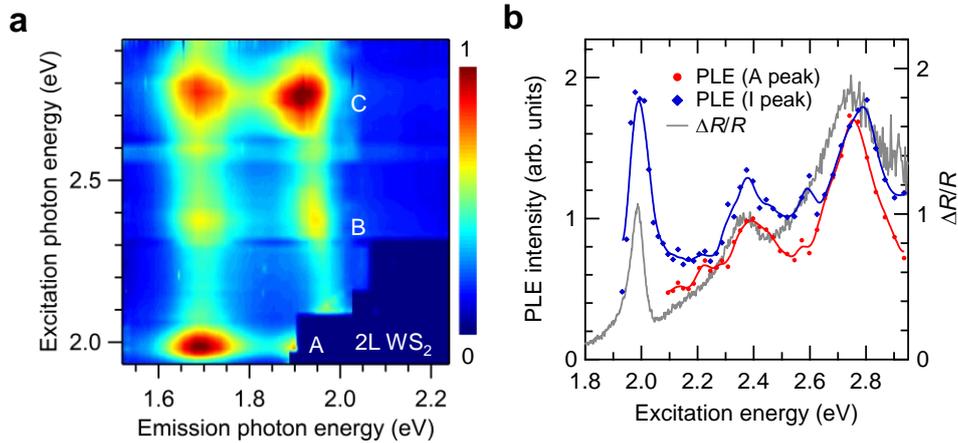

**Figure S4: PLE spectra of bilayer WS$_2$.** (**a**) PLE map and (**b**) PLE spectra for bilayer WS$_2$. The PLE intensity is evaluated as an integrated intensity of A and I peak in the PL spectra at each excitation energy. The optical contrast spectrum is also shown for comparison. The PLE of A peak is normalized by the B exciton peak of the optical contrast and the PLE of I peak is multiplied by the same factor as the PLE of A.

## S4. Lifetime of each process

We have summarized lifetimes for various processes of semiconducting transition metal dichalcogenides as shown in Table S1. A part of these lifetimes are considered for the interpretation of the excitation and relaxation pathways.

**Supplementary Table S1: Summary of lifetime for various processes**

| Ref. | Process | Life time | Material | Configuration |
|---|---|---|---|---|
| 8 | Hot exciton gas during the cooling process | 1-2 ps | CVD-grown $MoS_2$ on sapphire | Transient absorption spectroscopy with 3.16 eV excitation at 10 K by pump-probe (measuring the change in reflectivity or transmission) |
|  | Formation of biexciton | 2 ps |  |  |
|  | All excitons have relaxed to the lowest excited state (A exciton) | 4 ps |  |  |
| 9 | Polarization lifetime | 4.5 ps | Exfoliated monolayer $MoS_2$ on $SiO_2$/Si | Time-resolved PL spectroscopy with quasi-resonant excitation of the A-exciton transitions at 4 K-300 K |
|  | PL lifetime (non-radiative recombination of excitons) | 5 ps |  | Time-resolved PL exciting highly non-resonantly allowing both free carrier and exciton generation at 4 K |
|  | Localized exciton emission decay | 125 ps |  | Time-resolved PL spectroscopy with quasi-resonant excitation of the A-exciton transitions at 4 K |
| 10 | Intervalley (K-Γ) transfer time | 0.35 ps | Bulk $MoS_2$ crystal | Transient absorption spectroscopy pumping at the B or C (2.23 eV), probing at A (1.88 eV) peak at room temperature (measuring the change in |
|  | Energy relaxation time of | 50 ps |  |  |
|    | hot carriers | | | reflectivity) |
|    | Carrier lifetime | 180 ps ± 20ps | | |
| 11 | Exciton-exciton scattering | 500 fs | 4-5 L MoS$_2$ synthesized with a hot-wall furnace type system | Transient absorption spectroscopy pumping at B (2.04 eV) or C (3.10 eV), probing at A or B peak (1.70-2.21 eV) at room temperature (measuring the change in transmission experiments) |
| 12 | Defect assisted scattering from C to A | < 500 fs | Si$_3$N$_4$ substrate-suspended monolayer MoS$_2$ | Transient absorption spectroscopy pumping at 3.2 eV at room temperature (measuring the change in transmittance) |
|    | Carrier-phonon scattering from A to GS | 80 ps | | |
|    | Relaxation to trap states from A | 2-4 ps | | |
|    | Direct recombination from A to GS | 800 ps | | |
|    | Carrier-carrier scattering from C to A | 2 ps | Bulk MoS$_2$ crystal ~150 layers | |
|    | Carrier-phonon scattering from C to A | 20 ps | | |
|    | Intervalley scattering from A to indirect states | 2.6 ns | | |
|    | Indirect recombination to GS | > 2.6 ns | | |
| 13 | Exciton-exciton annihilation | < 50 ps | Exfoliated monolayer MoSe$_2$ on SiO$_2$/Si | Transient absorption spectroscopy pumping at 1.65 eV and probing at the A (1.51-1.57 eV) peak at room temperature (measuring the change in reflectivity) |
|    | Exciton lifetime | > 150 ps | | |
|    | Indirect bandgap transition (no exciton-exciton | 300-400 ps | Bulk monolayer | |
| | annihilation) | | | |
| 14 | Photocarrier recombination | 5 ps | Exfoliated monolayer $MoS_2$ on $SiO_2$/Si | Time-resolved PL spectroscopy at 4.5 K |
| | Exciton-phonon scattering | 70 ps | | Time-resolved PL spectroscopy at room temperature |
| 15 | Valley polarization decay time | 10 ps | CVD-grown monolayer $MoS_2$ | Transient absorption spectroscopy pumping at 1.88 eV and probing continuum pulses at 74 K (measuring change in transmittance) |
| 16 | Carrier lifetime | $100 \pm 10$ ps | Exfoliated few-layer $MoS_2$ on $SiO_2$/Si | Transient absorption spectroscopy pumping at the C (3.18 eV) and probing at A (1.88 eV) peak at room temperature (measuring change in reflection) |
| 17 | Phonon-related process | 7.1 ps, 61.3 ps | Exfoliated monolayer $MoS_2$ on $SiO_2$/Si | Transient absorption spectroscopy pumping at 1.96 eV and probing at 1.91 eV at 78 K (measuring change in reflection) |
| 6 | Exciton lifetime | > 50 ps | Exfoliated $MoS_2$ on hBN | Estimated from steady-state PL and quantum yields exciting with 1.96 eV continuous wave at 14 K |
| | Hole valley-spin lifetime | > 1 ns | | Polarization-resolved PL exciting with 1.96 eV continuous wave at 14 K |
| | Hole spin lifetime | Few hundreds of fs | Bilayer $MoS_2$ on hBN | |

**References**

1       Tongay, S. *et al.* Defects activated photoluminescence in two-dimensional semiconductors: interplay between bound, charged, and free excitons. *Sci. Rep.* **3**, 2657 (2013).

2       Tongay, S. *et al.* Broad-range modulation of light emission in two-dimensional semiconductors by molecular physisorption gating. *Nano Lett.* **13**, 2831-2836 (2013).

3       Mak, K. F. *et al.* Tightly bound trions in monolayer $MoS_2$. *Nat. Mater.* **12**, 207-211 (2013).

4       Mouri, S., Miyauchi, Y. & Matsuda, K. Tunable photoluminescence of monolayer $MoS_2$ via chemical doping. *Nano Lett.* **13**, 5944-5948 (2013).

5       Buscema, M., Steele, G. A., van der Zant, H. S. & Castellanos-Gomez, A. The effect of the substrate on the Raman and photoluminescence emission of single layer $MoS_2$. *arXiv preprint arXiv:1311.3869* (2013).

6       Mak, K. F., He, K., Shan, J. & Heinz, T. F. Control of valley polarization in monolayer $MoS_2$ by optical helicity. *Nat. Nanotechnol.* **7**, 494-498 (2012).

7       Kaasbjerg, K., Thygesen, K. S. & Jacobsen, K. W. Phonon-limited mobility in n-type single-layer $MoS_2$ from first principles. *Phys. Rev. B* **85**, 115317 (2012).

8       Sie, E. J., Lee, Y.-H., Frenzel, A. J., Kong, J. & Gedik, N. Biexciton formation in monolayer $MoS_2$ observed by transient absorption spectroscopy. *arXiv preprint arXiv:1312.2918* (2013).

9       Lagarde, D. *et al.* Carrier and polarization dynamics in monolayer $MoS_2$. *arXiv preprint arXiv:1308.0696* (2013).

10      Kumar, N., He, J., He, D., Wang, Y. & Zhao, H. Charge carrier dynamics in bulk $MoS_2$ crystal studied by transient absorption microscopy. *J. Appl. Phys.* **113**, 133702 (2013).

11      Sim, S. *et al.* Exciton dynamics in atomically thin $MoS_2$: Interexcitonic interaction and broadening kinetics. *Phys. Rev. B* **88**, 075434 (2013).

12      Shi, H. *et al.* Exciton dynamics in suspended mono layer and few-layer $MoS_2$ 2D crystals. *ACS Nano* **7**, 1072-1080 (2013).

13      Kumar, N. *et al.* Exciton-exciton annihilation in $MoSe_2$ monolayers. *arXiv preprint arXiv:1311.1079* (2013).

14  Korn, T., Heydrich, S., Hirmer, M., Schmutzler, J. & Schuller, C. Low-temperature photocarrier dynamics in monolayer MoS$_2$. *Appl. Phys. Lett.* **99**, 102109 (2011).

15  Mai, C. *et al.* Many body effects in valleytronics: Direct measurement of valley lifetimes in single layer MoS$_2$. *Nano Lett*. **14**, 202-206 (2013).

16  Wang, R. *et al.* Ultrafast and spatially resolved studies of charge carriers in atomically thin molybdenum disulfide. *Phys. Rev. B* **86**, 045406 (2012).

17  Wang, Q. *et al.* Valley carrier dynamics in monolayer molybdenum disulfide from helicity-resolved ultrafast pump–probe spectroscopy. *ACS Nano* **7**, 11087-11093 (2013).